# Intensity noise self-regulated solid-state laser at 1.5µm using an ASHG based Buffer Reservoir

**KEVIN AUDO,**[1,*] **AND MEHDI ALOUINI,**[1]

[1] *Institut FOTON, UMR CNRS 6082, Université Rennes 1, Campus Beaulieu, 35042 Rennes Cedex, France*

*\*kevin.audo@univ-rennes1.fr*

**Abstract:** An absorption mechanism based on second-harmonic generation (ASHG) is successfully implemented as a buffer reservoir in a solid-state Er,Yb:Glass laser emitting at telecom wavelength. We show that a slight ASHG rate conversion of 0.016% using a BBO crystal enables to cancel out the excess intensity noise at the relaxation oscillation frequency, i.e. 35 dB reduction, as well as to cancel the amplified spontaneous emission beating at the free spectral range resonances of the laser lying in the GHz range.

## 1. Introduction

Numerous applications require optical sources with low phase noise combined with low intensity noise such as lidar-radar [1], microwave photonics [2], atomic clocks [3] and metrology experiments [4]. Solid-state lasers are good candidates due to their typical thin linewidth and their capability to provide at the same time high optical power. Nevertheless, solid-state lasers suffer from excess intensity noise at low frequency, especially at the relaxation oscillation (RO) frequencies. The RO phenomenon is due to the class-B dynamics of solid-state lasers where the inversion population lifetime is higher than the photon lifetime inside the cavity leading to a resonant interaction between these two populations [5]. Different solutions have been investigated to reduce the intensity fluctuations at the RO frequency. The most common solution consists in using an electronic loop acting either on the pump laser [6] or on a voltage controlled attenuator [7]. Also, optical injection has been employed to reduce this intensity noise [8]. Besides, an efficient solution consists in changing significantly the laser behavior by turning it into a class-A laser. To reach this regime, the photon lifetime has to be higher than the inversion population lifetime leading a laser free from RO. This solution has been successfully implemented using a semiconductor active medium whose short carrier lifetime enables class-A operation with cm-long cavities [9]. However, due to the long population inversion lifetime in glass and crystal active media, reaching the class-A regime is not possible in solid-state lasers since it might require a kilometric long laser cavity while maintaining single longitudinal mode operation.

In parallel to these technics, we have explored an alternative approach relying on the introduction of a Buffer Reservoir (BR) inside the laser cavity. This approach has been proved to be powerful to reduce the excess intensity noise in solid-state lasers by drastically changing the dynamical behavior of the laser [10]. Indeed, by introducing a slight nonlinear absorption into the laser cavity, we have demonstrated the cancelation of intensity fluctuations at the RO frequency. This method offers the advantage to be passive, compact, easy to implement and extremely efficient in terms of noise reduction. Moreover, we have demonstrated that the required nonlinear absorption can be achieved by two-photon absorption mechanism (TPA) or by an absorption process induced by second harmonic generation (ASHG). It is worthwhile to notice that the action of the intracavity nonlinear absorption in the BR approach is strongly different from common uses of second-harmonic generation for frequency-doubling or for the squeezed light generation for example [11,12]. In BR approach, the nonlinear absorption is introduced in a very low efficiency in order to not disturb the laser operation.

Using ASHG as a BR presents a major advantage compared to TPA. From a technical point of view, TPA requires a semi-conductor whose gap energy is higher than the laser photon energy but also whose carrier recombination is very fast. Indeed, as presented in [10], the response time of the BR is limited by the carrier recombination lifetime of the semi-conductor and not by the TPA process itself. In other words, in order to reduce efficiently the excess intensity noise, the inverse of the carrier lifetime has to be much higher than the frequency at which the intensity fluctuations have to be reduced. The shorter this lifetime is, the more the BR will be efficient at higher frequencies. The bandwidth of the noise reduction is so defined by the inverse of the carrier recombination lifetime. These two requirements restrict the potential application of TPA to different regions of the optical spectrum. On the contrary, a BR based on an ASHG process enables to get rid inherently of the carrier recombination lifetime limitation as shown at 1.064 µm in a Nd:YAG laser [13]. Moreover, in regards of large number of non-linear crystals permitting ASHG, the BR approach can be potentially implemented irrespective to the operation wavelength. Thus, while the BR approach has been implemented at 1.55 µm using TPA [14], the possible use of ASHG as BR at this wavelength has never been explored.

In this paper, we show how sizing the appropriate crystal at 1.5 µm can lead to efficient noise cancellation in Er,Yb:Glass lasers.

## 2. Laser and measurement bench setup

We consider the laser presented in Fig. 1. It includes a phosphate glass active medium co-doped with Erbium and Ytterbium. The first side of the active medium is coated for high reflectivity at the laser wavelength (100% at 1.55 µm) and transmission of 95% at the pump wavelength. The active medium is pumped through a multimode laser diode providing 450 mW at 976 nm. The linear cavity is closed by a concave mirror with 5 cm radius of curvature and 0.5% transmission at 1.55 µm. A 1-mm-long YAG uncoated étalon is inserted inside the cavity to make the laser single mode.

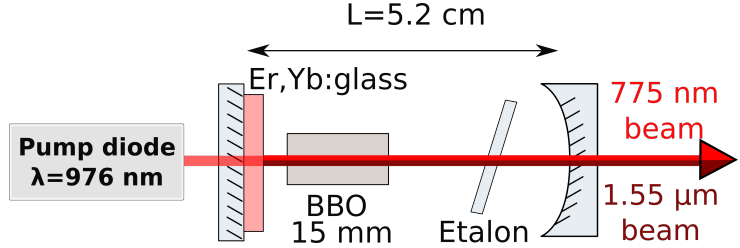

Fig. 1. Schematic representation of the Er,Yb:Glass laser. The intracavity slight ASHG is produced by a 15 mm long BBO crystal. A YAG étalon is inserted in the cavity to make the laser single mode

The ASHG process is realized through a 15 mm thick BBO crystal whose both faces are AR coated at 1.55 µm. The nonlinear crystal is cut with angles $\theta = 19.8°$ and $\varphi = 90°$ enabling phase matching for a type-I SHG at the laser wavelength 1.55 µm and at 775 nm, half the laser wavelength. It is worthwhile to notice that the conversion efficiency must be very low in our experiment in order to conserve the same static characteristic (threshold, power) of the laser as without BBO. This is ensured by first making the second-harmonic generated beam non-resonant in the cavity, the transmission of the output mirror being of 70% at 775 nm. Second, the laser beam in the BBO crystal is not focused and kept wide, i.e., around 160 µm beam diameter.

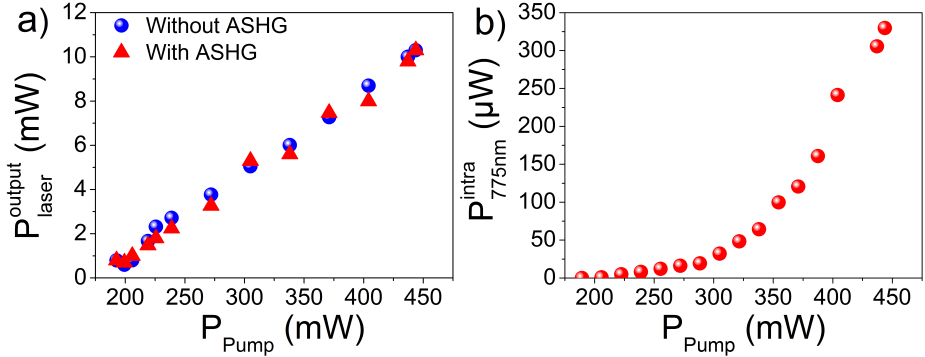

Fig. 2. a) Output power of the laser beam as a function of the pump power. The red and the blue dots correspond respectively to the case when the ASHG rate is null and when the ASHG rate is 0.016%. b) Optical power of the second harmonic beam as a function of the pump power.

We define the conversion efficiency as $\eta = P_{SHG}^{IN}/P_{IR}^{IN}$ where $P_{SHG}^{IN}$ and $P_{IR}^{IN}$ are respectively the intracavity power of the 775 nm and of the laser beams. When the two beams are phase matched, the maximum conversion efficiency is in our case 0.016% only for an output IR power of 10 mW. As reported in Fig. 2, with this slight ASHG, the average output power of the laser remains almost that of the laser without ASHG proving that the BR mechanism, when properly optimized, affects the noise characteristics only.

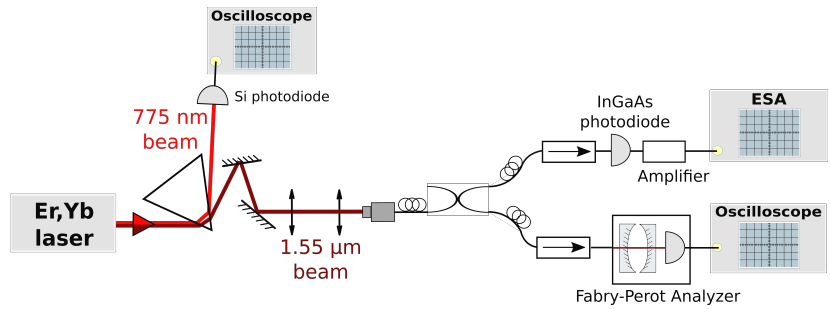

Fig. 3. Schematic of the measurement bench. The laser and the second harmonic beams are separated using a quartz prism. The power of the 775 nm beam is measured using a silicon photodiode. The laser beam is split into two arms. The first one includes a Fabry-Perot Analyzer whereas the second arm devoted to RIN measurements includes an InGaAs photodiode followed by a low noise amplifier and an electrical spectrum analyzer.

The intensity noise of the laser is characterized using the measurement bench depicted in Fig. 3. The fundamental and the second harmonic beams are separated using a quartz prism. The power of the frequency doubled beam is measured using a Silicon photodiode. The IR beam is split into two arms due to a 99%-1% coupler. The first arm including a Fabry-Perot Analyzer (FSR = 7 GHz) is devoted to spectral analysis in order to make sure that the laser is single mode and that no mode hopping occurs during the measurements. The second arm is dedicated to electrical noise measurements. The optical signal is measured with an InGaAs photodiode (bandwidth (BW) = DC-1 GHz) followed by a homemade low noise amplifier (Gain= 50 dB, BW= 1 kHz-500 MHz, NF= 1.4 dB). Finally, the noise spectra are recorded using an electrical spectrum analyzer from Rohde&Schwarz (BW: 10 Hz–3.6 GHz).

## 3. Results

First, let us focus on the resonant intensity noise at and around the RO frequency. We measure several RIN spectra for different conversion efficiencies (see Fig. 4). The conversion efficiency is changed by adjusting the phase matching when rotating the BBO crystal around the z- axis. It is important to keep in mind that for all ASHG rates, the optical power of the IR beam is almost constant, i.e. 10 mW, since the ASHG is made extremely low not to modify the static characteristic of the laser. The RIN spectra are measured for the same photocurrent: 1 mA.

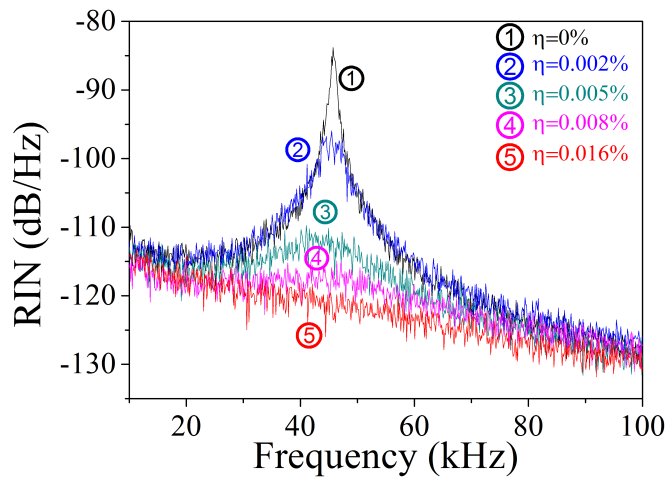

Fig. 4. RIN spectra of the Er,Yb:Glass laser recorded around the RO frequency for different ASHG efficiencies.

As one can see in Fig. 4, the RIN spectrum (1) without ASHG exhibits a peak at 42 kHz with an amplitude of -87 dB/Hz. This peak corresponds to the excess noise at the RO frequency. The BR is then introduced by increasing the ASHG efficiency through phase matching. One can see in Fig. 4 that the more we introduce the ASHG mechanism, the more the amplitude of the noise peak decreases. For an efficiency of 0.016% only, the noise at the RO frequency is completely cancelled. This corresponds to a reduction of at least 35 dB of the noise peak. Moreover, one can notice that the RO noise peak does not exhibit any frequency shift when the ASHG efficiency is increased. This proves that the BR based on ASHG can be considered as instantaneous compared to that obtained with TPA which is limited by the carrier recombination lifetime [10].

Let us now focus on the effect of the ASHG on the excess noise of the laser appearing at higher frequencies. In particular, we focus on the peaks appearing at the harmonics of the free spectral range (FSR) of the cavity, i.e. in the GHz range. These peaks correspond to the beat-notes between the oscillating mode and the amplified spontaneous emission present in the longitudinal non-lasing adjacent modes which are collected by the photodiode. In order to cancel this excess noise, the BR lifetime has to be shorter than the inverse of the frequencies of the beat-notes peaks. Using ASHG, this condition is obviously fulfilled [13].

For that study, we replace the photodiode by an InGaAs photodiode with DC-16 GHz followed by a low noise amplifier (Gain = 47 dB, BW= 100 MHz-20 GHz, NF= 2.7 dB). The electrical spectrum analyzer is replaced by an Aeroflex 3250 series (BW: 1 kHz–26.5 GHz). The photocurrent is increased to 1.9 mA in order to reduce the measurement noise floor to the shot noise limit making the peaks more visible.

As shown by the plots labeled 1 in Fig. 5(a) and Fig. 5(b), without the BR, the RIN spectra of the laser present noise peaks at 2.06 GHz and 4.11 GHz, corresponding respectively to once and twice the FSR of the cavity. The amplitude of these peaks is -152 dB/Hz. The next beat-note peaks at higher frequencies are below the shot noise level due to the filtering of the intracavity étalon. Then, we insert the ASHG in the laser with a conversion efficiency of 0.016% by positioning the crystal in the phase matching condition. As presented by the RIN spectra labeled 2 in Fig. 5(a) and Fig. 5(b), the insertion of the ASHG, even very slight, enables to significantly reduce the excess noise by 6 dB at least. The noise of the laser is thus limited by the shot noise level over the whole spectrum from 300 MHz up to 16 GHz, i.e. -157.3 dB/Hz for a photocurrent of 1.9 mA.

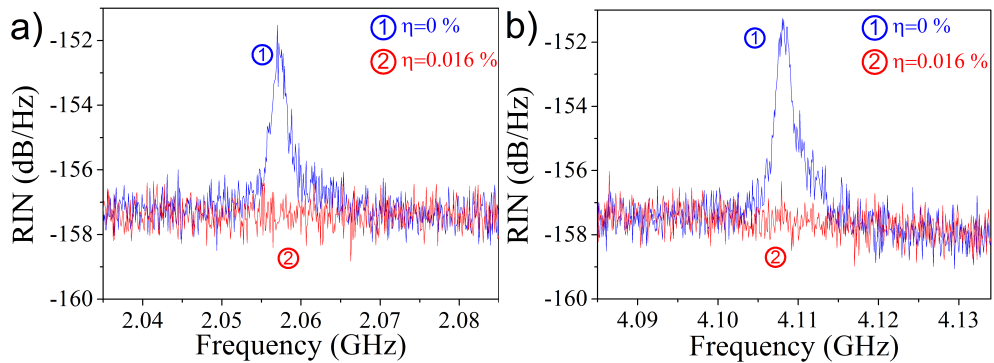

Fig. 5. RIN spectra recorded around frequencies corresponding to a) the first FSR frequency and b) the second FSR frequency. The blue spectra labeled (1) are those obtained without ASHG and the red ones labeled (2) are those obtained with ASHG.

## 4. Conclusion

In this paper, we have demonstrated the cancellation of the excess intensity noise in a Er,Yb:Glass laser operating at 1.55 µm by implementing a Buffer Reservoir based on ASHG. The ASHG is obtained using a BBO crystal. A reduction of at least 35 dB of the resonant noise at the RO frequency is obtained with a conversion efficiency as low as 0.016% enabling to maintain the static characteristics of the laser unchanged. Moreover, we have shown that the excess noise due to the beat-note between amplified spontaneous emission and the lasing mode appearing at high frequencies is also cancelled. The realization of a very low noise laser operating at telecom wavelength open new perspectives for applications requiring very low noise laser sources at 1.55 µm with power scalable to several Watt such as in coherent telecom links, microwave photonics or lidar systems.

## 5. Acknowledgments

This work is supported by the Direction Générale de l'Armement (DGA), Région Bretagne and Thales Research and Technology.